\documentclass{iau}
\usepackage{graphicx}

\title[Magnetic fields around Bok Globules] 
{Magnetic field dispersion in the \\neighbourhood of Bok Globules}

\author[C. V. Rodrigues et al.]   
{C.~V. Rodrigues$^1$, V. de S. Magalh\~aes$^1$, J.~W. Vilas-Boas$^1$, G. Racca$^1$ \and  A. Pereyra$^1$}

\affiliation{$^1$Divis\~ao de Astrof\'\i sica, Instituto Nacional de Pesquisas Espaciais\\
Av. dos Astronautas, 1758 -- S\~ao Jos\'e dos Campos -- SP -- Brazil\\ email: {\tt 
claudiavr@das.inpe.br}}

\pubyear{2013}
\volume{302}  
\jname{Magnetic Fields Throughout Stellar Evolution}
\editors{A.C. Editor, B.D. Editor \& C.E. Editor, eds.}
\begin{document}

\maketitle

\begin{abstract}
We performed an observational study of the relation between the interstellar magnetic field alignment and star formation in twenty (20) sky regions containing Bok Globules. The presence of young stellar objects in the globules is verified by a search of infrared sources with spectral energy distribution compatible with a pre main-sequence star. The interstellar magnetic field direction is mapped using optical polarimetry. These maps are used to estimate the dispersion of the interstellar magnetic field direction in each region from a Gaussian fit, $\sigma_B$. In addition to the Gaussian dispersion, we propose a new parameter, $\eta$, to measure the magnetic field alignment that does not rely on any function fitting. Statistical tests show that 
the dispersion of the magnetic field direction
is different in star forming globules relative to quiescent globules.
Specifically, the less organised magnetic fields occur in regions having young stellar objects. 
\end{abstract}
 
\firstsection 
              
\section{Introduction}

The star formation results from the interplay between physical ingredients as the gravitational field, gas pressure, magnetic fields, and turbulence (e.g., \cite{crutcher2012}). In particular, the importance of magnetic fields and turbulence is still an open issue. Other unanswered question is the origin of the interstellar turbulence.

There are some indicatives that star forming regions have less organised magnetic fields than quiescent regions. \cite{pereyra2000} studied the magnetic field direction along Musca Dark Cloud and found that there is an increase of the dispersion of the magnetic field direction (DMFD) near a young stellar object. The same happens in Pipe Nebula (\cite{alves2008}; \cite{franco2010}): the B59 region has the largest DMFD in this dark cloud. Additionally, the DMFD in the interstellar medium near Herbig-Haro objects is qualitatively consistent with that of the B59 region (\cite{targon2011}).

Bok Globules are among the simplest interstellar regions. Hence they are appropriate to study the role of different physical aspects on the star formation. Do the Bok Globule properties differ as a function of the presence of star formation? To help to answer this question, we performed an observational project to verify whether the magnetic field organisation differs as a function of the presence of star formation in a sample of Bok Globules. We used optical polarisation to map the magnetic field direction. This study is partially presented in \cite{victor}.

\section{The Bok Globules sample and their association with star formation}

The Bok Globules sample used in this study is the same presented in \cite{racca2009} (see Tab.~\ref{tab}). These globules are not associated with bright nebulae or molecular complexes. 

\cite[Racca et al. (2009)]{racca2009} classified 10 globules as having star formation by the presence of an associated IRAS source. In the present study, we revise this classification based on 
new infrared surveys. 
Data from WISE (\cite{wise}), 2MASS (\cite{2mass}), AKARI (\cite{akari}), and GLIMPSE (\cite{glimpse1}; \cite{glimpse2}) were used to obtain the spectral energy distribution (SED) for the identified sources. These SEDs were fitted using Robitaille et al. (2006, 2007)'s  models to check if they are consistent with young stellar objects or reddened foreground/background objects. Table~\ref{tab} presents the resulting classification.

\setlength{\tabcolsep}{10pt}

\begin{table}
\begin{center}
\caption{Estimates of the alignment degree of the magnetic field direction around our Bok globules  sample. The table is separated in globules with and without star formation. }
\label{tab}

\begin{tabular}{cc}
\begin{minipage}[t]{6cm}
\vspace{-5.11cm}
\begin{center}
\begin{tabular}{ccc}
 \hline 
 \multicolumn{3}{c}{Star forming region} \\
\hline
Globule	& $\eta$ & $\sigma_B$ \\
\hline
BHR~016	& 	0.71	& 13.4 \\
BHR~044	& 	0.17	& -  \\
BHR~053	& 	0.68	& 26.8 \\
BHR~058	& 	0.88	& 11.3 \\
BHR~075	& 	0.87	& 13.4 \\
BHR~117	& 	0.90	& 12.0 \\
BHR~138	& 	0.98	& 5.2 \\
BHR~139	& 	0.99	& 5.6 \\
BHR~140	& 	0.94	& 7.9 \\
BHR~148-151	& 	0.67	& -  \\
BHR~059	& 	0.51	& - \\
BHR~145	& 	0.44	& - \\
BHR~113	& 	0.99	& 5.1  \\ \hline
\end{tabular}
\end{center}

\end{minipage} &

\begin{minipage}[b]{6cm}
\begin{center}

\begin{tabular}{ccc}
 \hline 
 \multicolumn{3}{c}{No star formation} \\
 \hline
Globule	& $\eta$ & $\sigma_B$ \\
\hline
BHR~034 & 	0.92	& 13.8 \\
BHR~074	& 	0.91	& 10.6 \\
BHR~121	& 	0.97	& 7.1 \\
BHR~126	& 	0.93	& 11.2 \\
BHR~133	& 	0.91	& 13.8 \\
BHR~144	& 	0.73	& 15.1 \\
BHR~111$^{1}$	& 	0.94	& 10.2 \\
\hline
\end{tabular}

$^1$BHR~111 is associated with a Very Low Luminosity Object (\cite{vello}).
\end{center}

\end{minipage}

\end{tabular}
\end{center}
\end{table}

\section{Optical polarimetry and the alignment of the magnetic field}

\begin{figure}[bth]
\begin{center}
\begin{tabular}{cc}

\begin{minipage}{6cm}
\includegraphics[width=5cm]{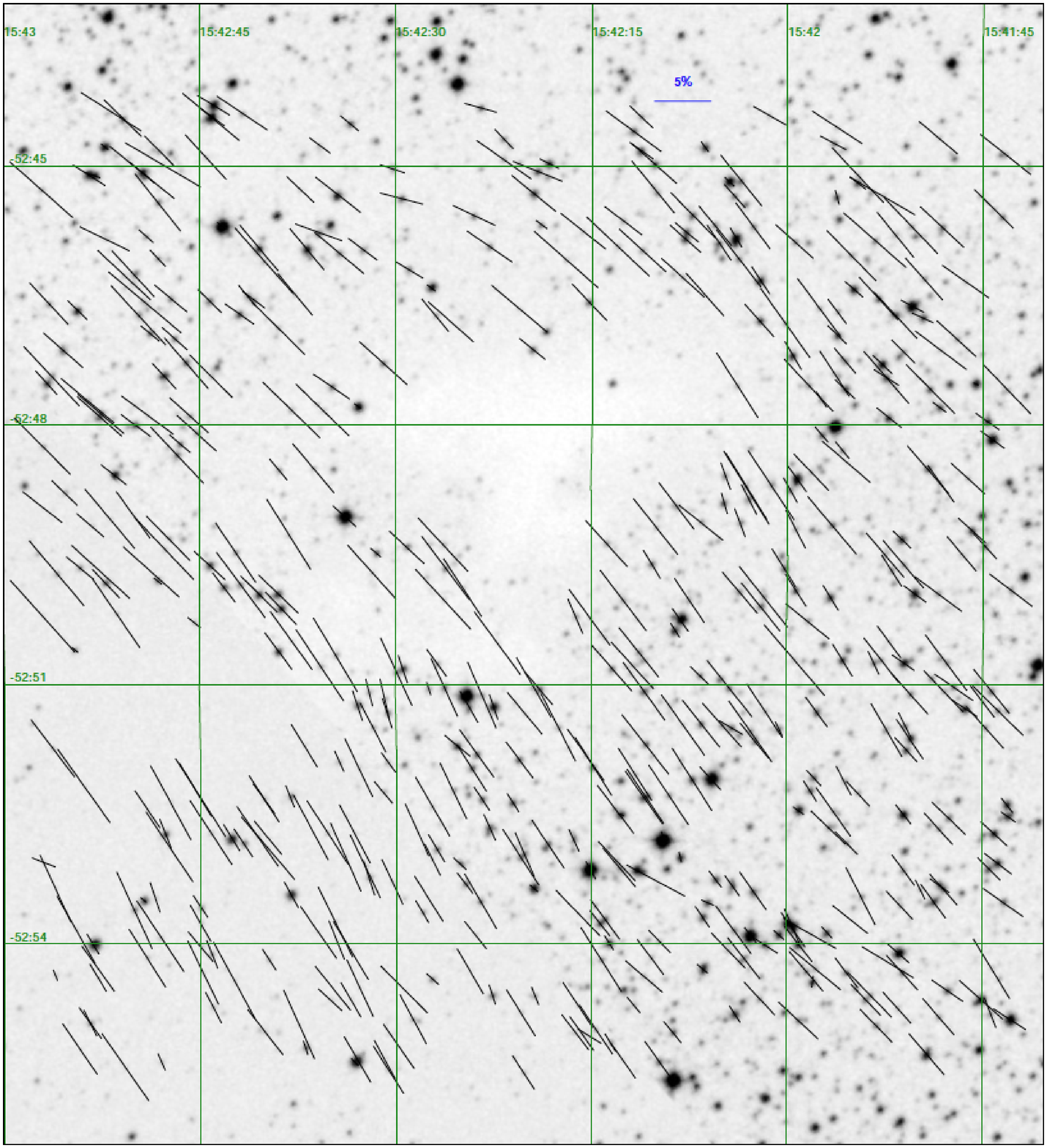}
\end{minipage} &

\begin{minipage}{6cm}
\hspace{-1cm}
\includegraphics[width=5cm,angle=-90]{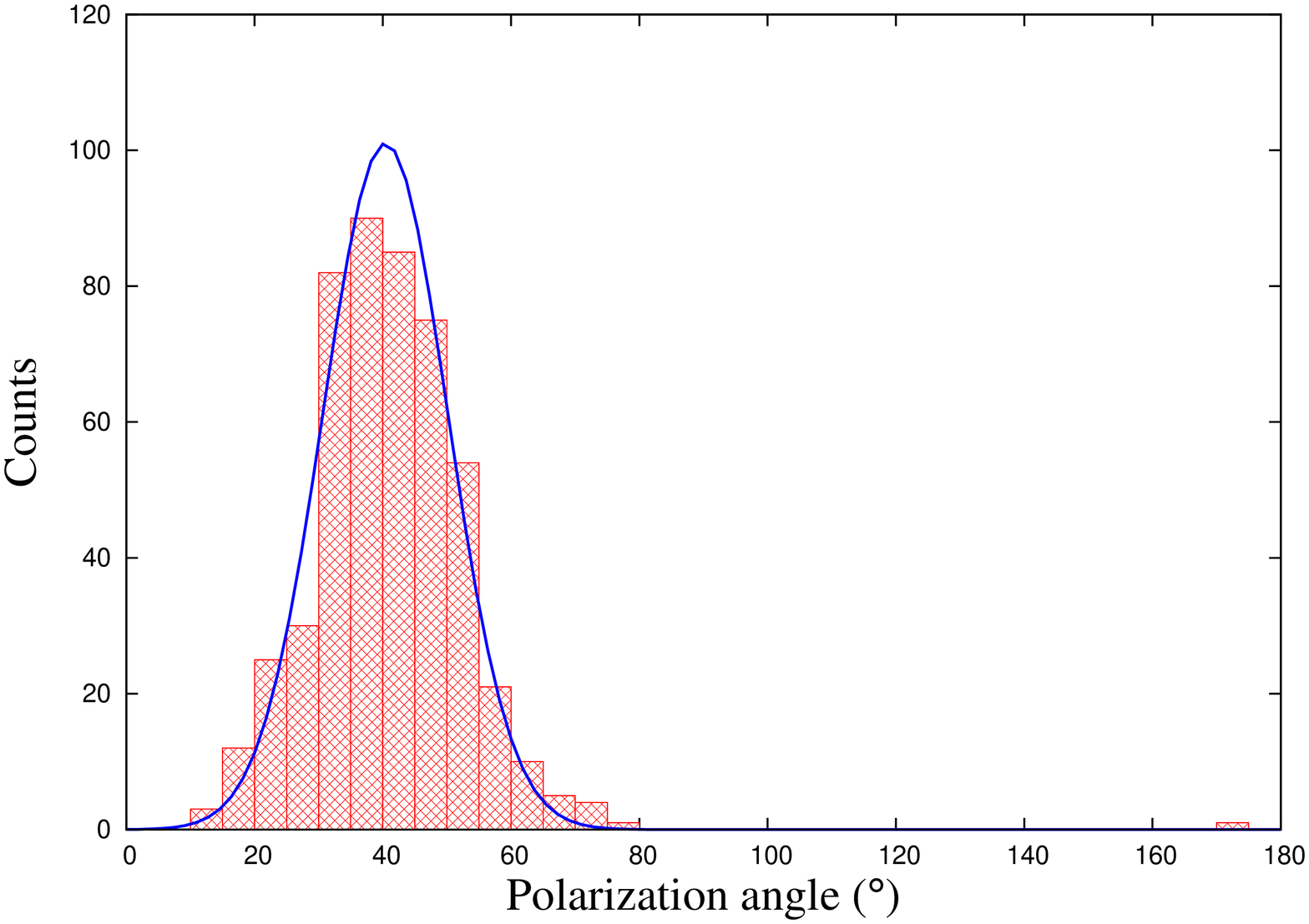}
\end{minipage}

\end{tabular}
\end{center}
\caption{(Left) Optical polarimetry in the line of sight of BHR~111 in the I band superposed on a DSS2 red image. (Right) The corresponding histogram of the linear polarisation angle. 
The superposed line is a Gaussian fit.
}
\label{bhr111}
\end{figure}

The interstellar magnetic field was mapped using optical polarimetry. The data were collected at the 0.6m telescope of the \textit{Observat\'orio do Pico dos Dias}\footnote{\noindent Managed by  \textit{Laborat\'orio Nacional de Astrof\'\i sica}/Brazil.}. We used a CCD polarimeter \cite{magalhaes1996} and an $I_C$ filter. This instrumental configuration provides polarimetry of point sources in a field-of-view of 11 x 11 $arcmin^2$ (e.g., Fig.~\ref{bhr111}.left).

 \begin{figure}[bth]
\begin{center}
\begin{tabular}{cc}

\begin{minipage}{6cm}
\includegraphics[width=5cm]{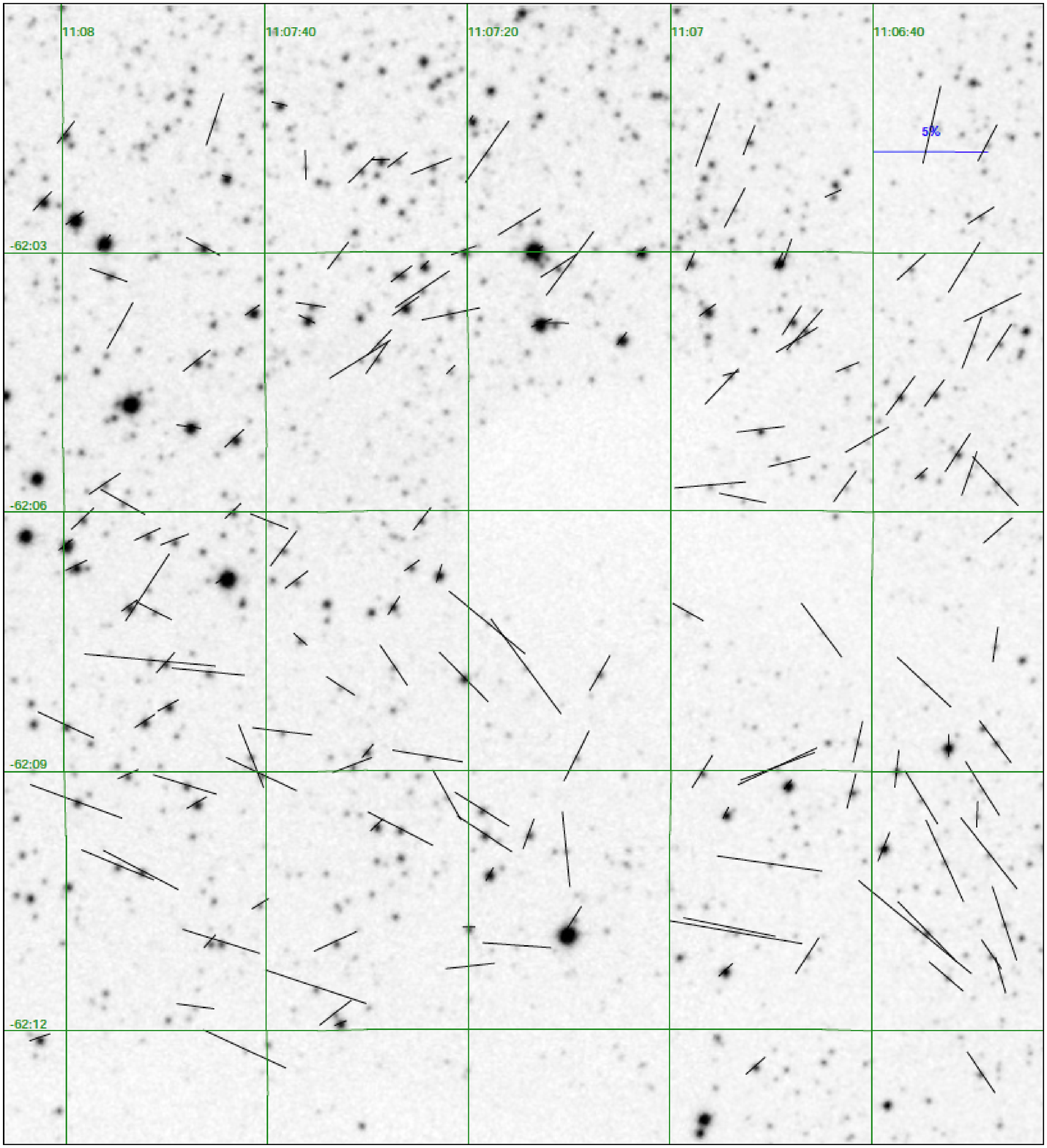}
\end{minipage} &

\begin{minipage}{6cm}
\hspace{-1cm}
\includegraphics[width=5cm,angle=-90]{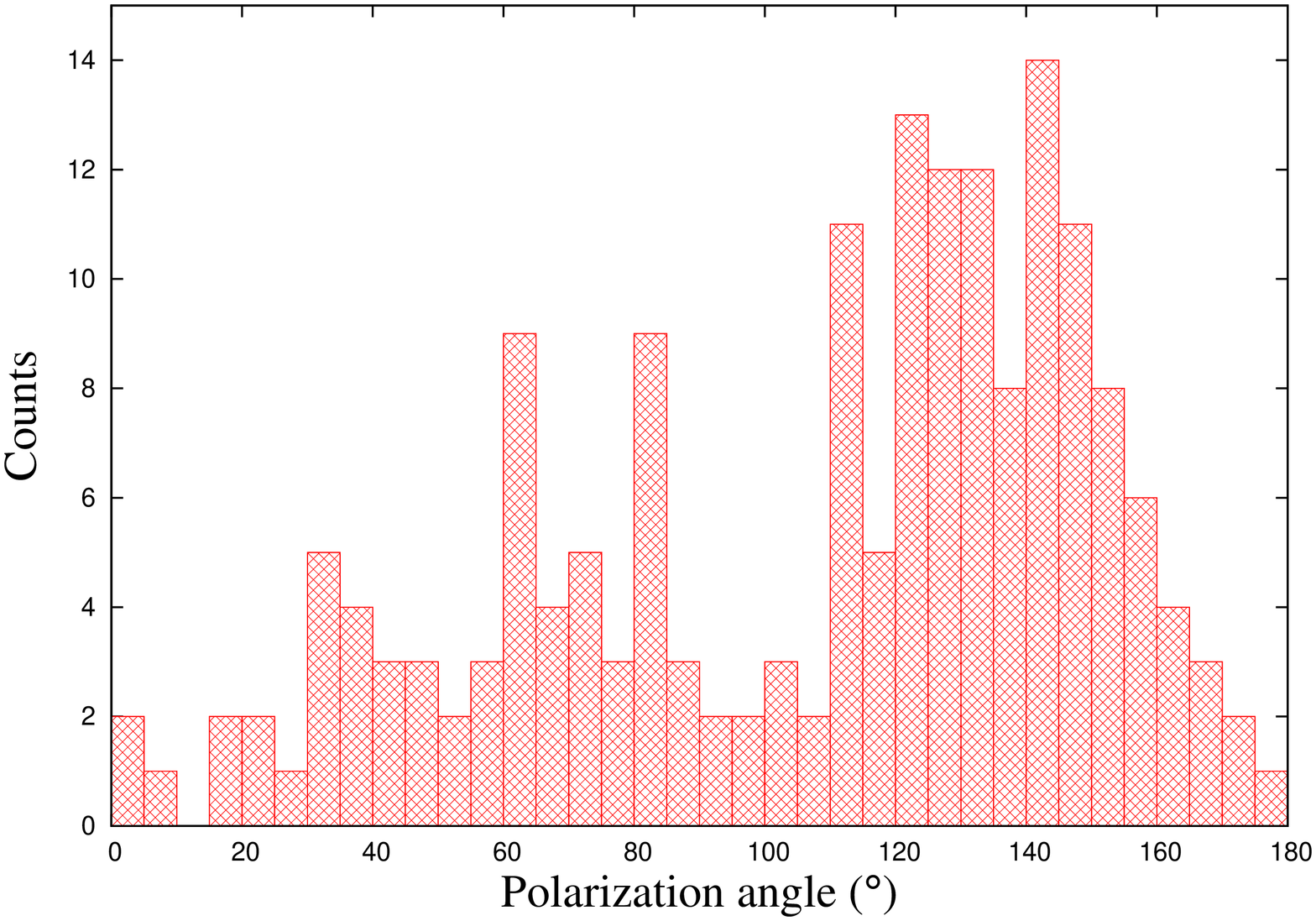}
\end{minipage}

\end{tabular}
\end{center}
\caption{As Figure~\ref{bhr111}, for BHR~059.}
\label{bhr059}
\end{figure}
 
Usually the alignment degree of the interstellar magnetic field direction is quantified by the Gaussian dispersion of the polarisation angle, $\sigma_B$ (e.g., Fig. \ref{bhr111}.right). However, a Gaussian fit is not always possible. An example is the BHR~059 field (Fig. \ref{bhr059}). To let a quantitative analysis for all fields, we propose a new estimator of the magnetic field alignment, $\eta$:

\begin{equation}
\eta = \frac{\overline{P}}{\overline{|P|}}, 
\end{equation}

\noindent where $\overline{P}$ is the vectorial average of the polarisation vectors and $\overline{|P|}$ is the arithmetic average of the polarisation module of the field stars. If all objects have the same polarisation angle, $\eta = 1$; if the polarisation angles are randomly distributed, $\eta$ tends to zero. Table~\ref{tab} presents $\eta$ and $\sigma_B$  for the observed fields. We could not estimate $\sigma_B$ for BHR~059 and other fields presenting a non Gaussian distribution of polarisation vectors. 

\section{Results and discussion}

\begin{figure}[t]
\begin{center}
\begin{tabular}{cc}

\begin{minipage}[t]{4cm}
\caption{Polarisation alignment efficiency, $\eta$, and DMFD, $\sigma_B$, for star forming globules (red triangles) or quiescent globules (blue circles).}
\label{fig_eta_sigma}
\end{minipage}
&

\begin{minipage}{7.5cm}
\includegraphics[width=4.5cm,angle=-90]{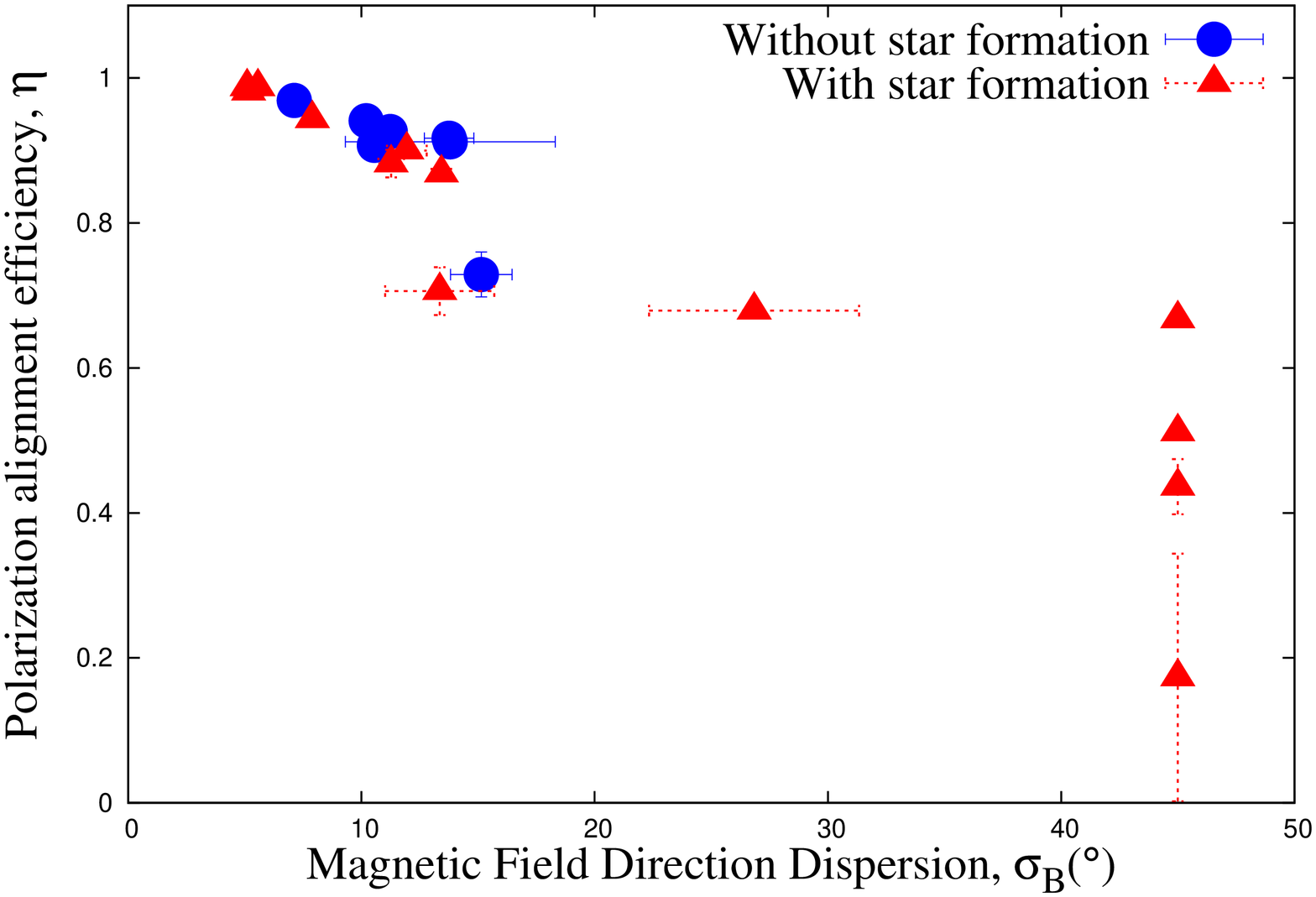}
\linethickness{0.7pt} 
\put (-149,-40)  {\rotatebox{-30}  {\vector(1,0){130}}} 
\put (-132,-59) {\rotatebox{-30}  {{\scriptsize Less organized fields}}}
\end{minipage}

\end{tabular}
\end{center}
\end{figure}

Figure~\ref{fig_eta_sigma} shows graphically $\eta$ and $\sigma_B$ for our sample. The magnetic field organization decreases from the left upper region to the right lower corner. Organised magnetic fields are found in regions with or without star formation. However, the less organised fields are only found for regions presenting star formation.

Using the Kolmogorov-Smirnov test, we found that the probability that the globules with star formation have the same $\eta$ ($\sigma_B$) distribution of quiescent globules is 8\% (68\%). 


These results suggest that the alignment of the interstellar magnetic field is different in regions with star formation compared with quiescent regions, and less organised fields are found in star forming regions. 
A possible interpretation is that the star formation injects kinetic energy in the interstellar medium, increasing the turbulence and disorganising the magnetic field; the organisation of the magnetic field lines is related to the gas turbulence according to \cite{cf1953}, considering energy equipartition. Another possibility is that the star formation is favoured in more turbulent regions. 


\begin{figure}[hbt]
\begin{center}
\includegraphics[angle=-90,width=6cm,trim=0cm 0cm 0 0 ]{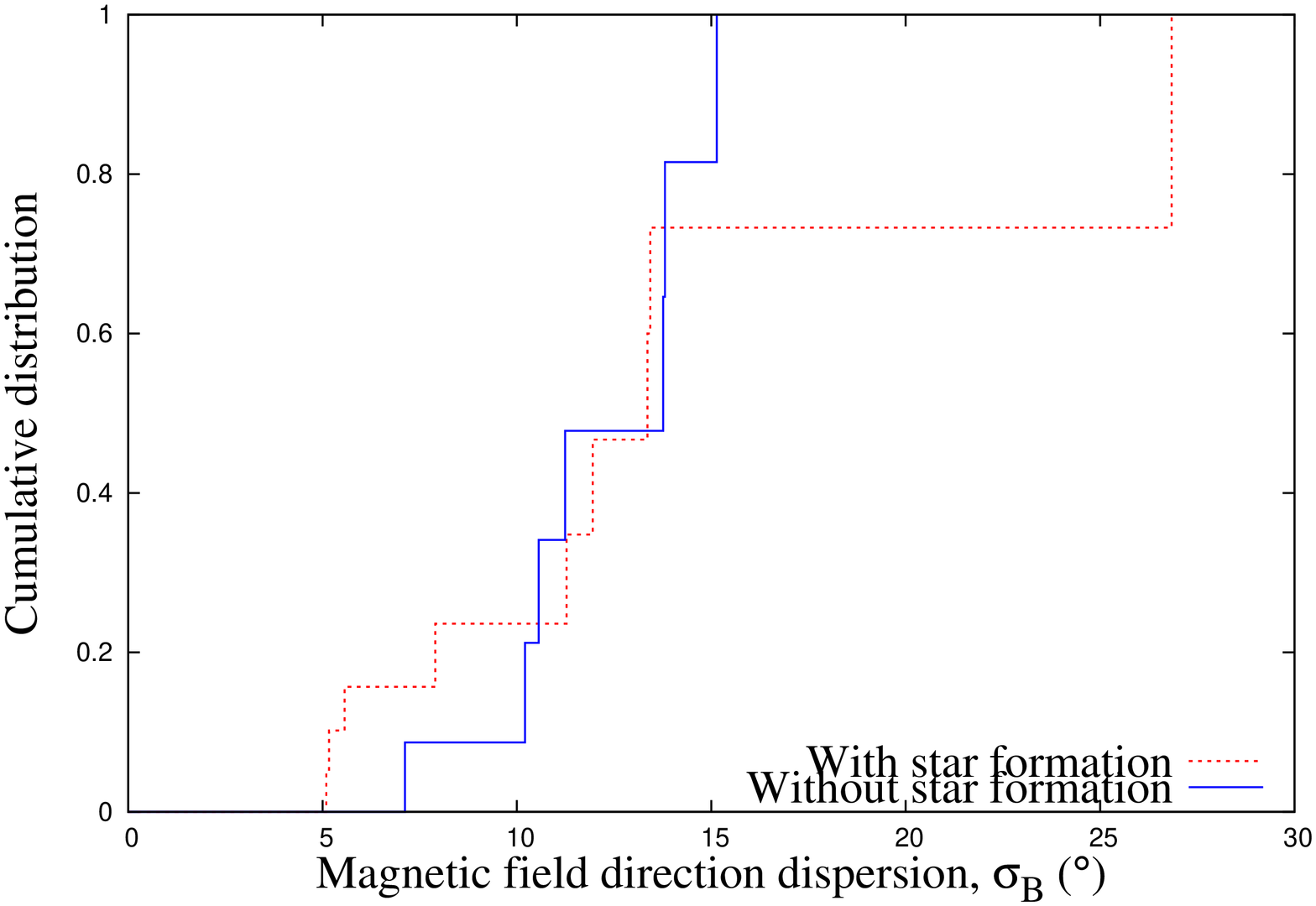}
\includegraphics[angle=-90,width=6cm,trim=0cm 0cm 0 0 ]{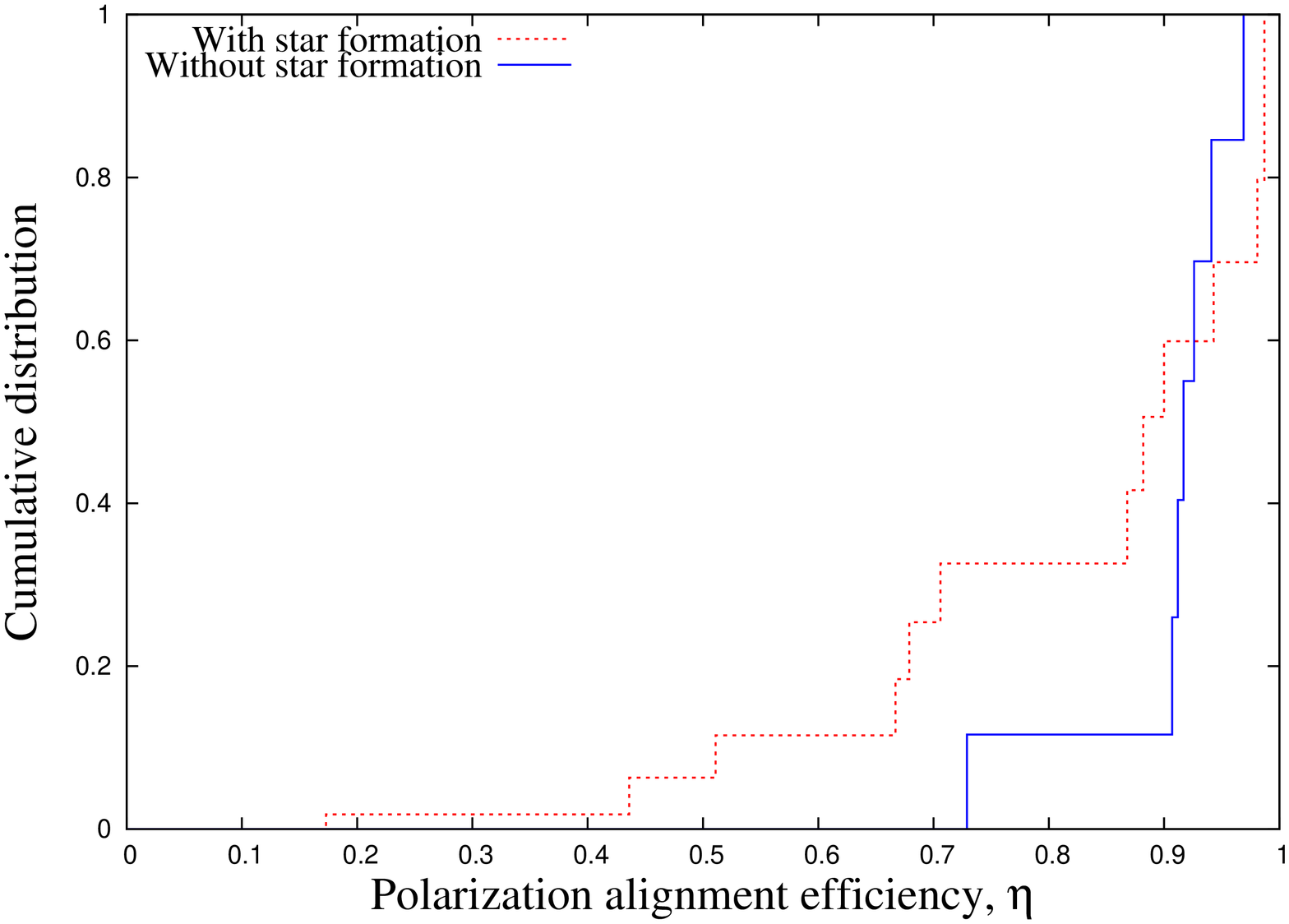}
\caption{(Left) Cumulative distribution of $ \sigma_b$ for globules having star formation (red dotted line) or quiescent globules (blue solid line). (Right) The same for $\eta$.}
\label{fig_cumul}
\end{center}
\end{figure}

\bigskip

\textbf{Acknowledgements:} CVR thanks Grant 2010/01584-8, S\~ao Paulo Research Foundation (FAPESP). This research makes use of: data products from WISE and 2MASS; the NASA/IPAC Infrared Science Archive;  observations made with AKARI and the Spitzer Space Telescope;  the SIMBAD database; the NASAÕs ADS Service; the NASA's SkyView facility; Aladin; and DSS.

\end{document}